# Emergence of Topologically Non-trivial Spin-polarized States in a Segmented Linear Chain

Thang Pham[1,2,3,4,†], Sehoon Oh[1,2,†], Scott Stonemeyer[1,2,4,5], Brian Shevitski[1,2], Jeffrey D. Cain[1,2,4], Chengyu Song[6], Peter Ercius[6], Marvin L. Cohen[1,2], and Alex Zettl[1,2,4*]

[1]*Department of Physics, University of California at Berkeley, Berkeley, CA 94720, USA*

[2]*Materials Sciences Division, Lawrence Berkeley National Laboratory, Berkeley, CA 94720, USA*

[3]*Department of Materials Science and Engineering, University of California at Berkeley, Berkeley, CA 94720, USA*

[4]*Kavli Energy NanoSciences Institute at the University of California at Berkeley, Berkeley, CA 94720, USA*

[5]*Department of Chemistry, University of California at Berkeley, Berkeley, CA 94720, USA*

[6]*National Center for Electron Microscopy, The Molecular Foundry, One Cyclotron Road, Berkeley, CA 94720, USA*

†These authors contributed equally

*e-mail: azettl@berkeley.edu

**The synthesis of new materials with novel or useful properties is one of the most important drivers in the fields of condensed matter physics and materials science. Discoveries of this kind are especially significant when they point to promising future basic research and applications. Van der Waals bonded materials comprised of lower-dimensional building blocks have been shown to exhibit emergent properties when isolated in an atomically thin form[1-8]. Here, we report the discovery of a transition metal chalcogenide in a heretofore unknown segmented linear chain form, where basic building blocks each consisting of two hafnium atoms and nine tellurium atoms ($Hf_2Te_9$) are van der Waals bonded end-to-end. First-principle calculations based on density functional**

theory reveal striking crystal-symmetry-related features in the electronic structure of the segmented chain, including giant spin splitting and nontrivial topological phases of selected energy band states. Atomic-resolution scanning transmission electron microscopy reveals single segmented $Hf_2Te_9$ chains isolated within the hollow cores of carbon nanotubes, with a structure consistent with theoretical predictions. Van der Waals-bonded segmented linear chain transition metal chalcogenide materials could open up new opportunities in low-dimensional, gate-tunable, magnetic and topological crystalline systems.

**Main**

The isolation of single layers of vdW bonded sheet-like materials has launched a new era of two-dimensional (2D) materials exploration, with the discovery of intriguing physical phenomena such as the fractional quantum Hall effect in graphene[1] and an indirect-to-direct bandgap transition for monolayers of the transition metal dichalcogenide $MoS_2$[2]. Recently, chain-like transition metal trichalcogenides (TMTs) have been examined at the one-dimensional (1D) limit inside carbon or boron nitride nanotubes, with, for example, charge-induced torsional waves appearing in single-chain $NbSe_3$[3].

Here, we investigate the transition metal chalcogenide (TMC) hafnium telluride (Hf-Te). Bulk $HfTe_2$ is a semimetal comprised of 2D layers with vdW bonds between layers[7,9], while bulk $HfTe_3$ is a metal composed of 1D trigonal prismatic chains with vdW bonds between the chains[10,11]. Isolation of a monolayer of $HfTe_2$ shows flattening of the valence band[7], while encapsulation of chains of $HfTe_3$ within carbon nanotubes (CNTs) results in a size-driven metal-insulator transition as the chain number passes from 4 to 3[4]. A single chain of $HfTe_3$ encapsulated within a CNT supports a short-wavelength trigonal antiprismatic rocking distortion that drives a significant energy gap, but the chain is otherwise structurally robust[4].

We show here that an entirely new low-dimensional form of Hf-Te is possible. The new form consists of structurally coherent zero-dimensional "blocks", each comprised of two Hf atoms and nine Te atoms (i.e. $Hf_2Te_9$), arranged end-to-end in a *segmented linear chain*, with vdW bonds linking the blocks or chain segments. The segmented chain resides inside the hollow core of a CNT, which protects the chain from oxidation and facilitates experimental characterization. First principles calculations reveal an especially rich electronic structure for the segmented chain. The bands near the chemical potential are fully spin-polarized in one direction and *k*-dependent giant spin splitting (GSS) of the bands occurs due to one mirror symmetry which is broken by crystal momentum *k*, and another mirror symmetry which is preserved for all *k*. Calculated Zak phases of the bands suggest externally tunable topological invariance.

**Structural Characterization**

Segmented chain $Hf_2Te_9$ specimens are synthesized using a method similar to that as described previously for the preparation of few-chain $HfTe_3$[4]. Briefly, end-opened multiwall carbon nanotubes are reacted together with Hf powder and Te shot, with iodine ($I_2$) as a transport agent, in a sealed ampule at 575 °C (see Methods for details).

Fig. 1 shows structural characterization of a segmented $Hf_2Te_9$ chain encapsulated within a CNT. Fig. 1a shows an atomic-resolution high-angle annular dark-field (HAADF) scanning transmission electron microscopy (STEM) image of the chain. The inner diameter of the CNT host (not easily visible in Fig. 1a) is 1.1 nm. The chain is clearly and strikingly segmented into regularly spaced blocks, each with distinct substructure. Almost all (>95%) of the blocks imaged contain two bright spots in the center, suggestive of transition metal atoms, surrounded by a handful of dimmer spots, suggestive of chalcogen atoms.

To aid in the identification of the precise atomic structure of the blocks, we perform density functional theory (DFT) calculations. Several candidate structures of the segmented chain with various chemical formulae isolated in a vacuum region of 50 Å × 50 Å perpendicular to the chain are constructed, and the atomic positions of the candidate structures are fully relaxed by minimizing the total energy. Using the relaxed atomic positions of the segmented chain isolated in vacuum and those of a pristine (8,8) CNT (with inner diameter 1.1 nm), we construct the atomic structure of the segmented chain inside the (8,8) CNT. The atomic positions of the segmented chain are relaxed further by minimizing the total energy while those of the CNT are fixed. The encapsulation does not alter the atomic structure of the segmented chain appreciably. Among all of the candidate structures investigated, only one relaxed structure, with a chemical formula of $Hf_2Te_9$, matches well with the experimental data.

Figs. 1b,c show the theoretically predicted structure of the segmented chain encapsulated inside the (8,8) CNT. Fig. 1b shows the segmented chain together with the CNT, while Fig. 1c shows the detailed structure of a single block (see Extended Data Fig. 1 as well). The length of a block is 7.89 Å, and blocks are separated by 3.65 Å. The block atomic structure has two mirror symmetries and the mirror planes perpendicular to the $y$- and $z$-axis are respectively denoted as $M_y$ and $M_z$ in Fig. 1c. Each block has two Hf atoms and nine Te atoms, with sequence $Te_3$-Hf-$Te_3$-Hf-$Te_3$ along the chain ($z$) direction. As shown additionally in Extended Data Fig. 1, the adjacent three Te atoms form isosceles triangles. The isosceles triangle at the center of the block is on the mirror plane, $M_z$, perpendicular to the chain direction, while those on the sides of the block cant toward the center of the block. Each Hf atom is surrounded by six Te atoms forming a trigonal prismatic structure (Extended Data Fig. 1), notably unlike the single-chain $HfTe_3$, which prefers a trigonal antiprismatic form[4]. (More details about the atomic structure models and symmetry are presented in Extended Data Fig. 1.)

To further validate the proposed block atomic structure, we simulate a block STEM image, as shown in Fig. 1d. To compare this prediction to experimental STEM data, we first generate a composite STEM image by averaging 227 orientationally-similar experimental STEM single-block images, as shown in Fig. 1e. Comparison of Figs. 1d and 1e, as well as the dimensions of a block and the separation gap, shows good agreement between theory and experiment. We have also performed electron energy loss spectroscopy (EELS) on segmented chain samples (see Extended Data). The Te M-edges of the EELS spectra are consistent with the atomic structure shown in Fig. 1c. We note that previous studies have shown that CNTs can be filled by pure metals and pure chalcogens. We have performed control experiments and found no evidence for segmented chain structure within CNTs using only Hf, only Te, or only $I_2$ as feedstock.

**Electronic Structure**

We investigate the electronic structure of the segmented $Hf_2Te_9$ chain by DFT calculations. Because of the presence of the mirror symmetry with respect to $M_y$ plane (Fig. 1c and Extended Data Fig. 1b), which is not broken by the crystal momentum $k$, the $x$- and $z$-components of spin and orbital angular momentum (OAM) with respect to the atomic positions on the plane should be zero regardless of $k$[12]. However, the crystal momentum $k$ breaks the mirror symmetry with respect to the $M_z$ plane, and the energy eigenstates are not the eigenstates of $M_z$, so the $y$-component of the OAM of the states can have finite value except for the time reversal invariant momentum (TRIM) points, the center, and the edge of the Brillouin zone ($k = 0, \pi/a$, where $a$ is the unit cell length).

Figs. 2a,d show the electronic band structure of the segmented chain isolated in vacuum obtained without considering spin-orbit interaction (SOI). The bands in Fig. 2d are colored with the expectation values of the $y$-component of local OAM of the states $|\psi_{nk}\rangle$, with the band index $n$ and the crystal momentum $k$, integrated over all the atoms in the unit cell, $<L_y> = \sum_{i \in cell} <L_y^i>$, where $<L_y^i>$ is the expectation value of the $y$-component of local OAM with respect to the $i$-th atom of the states $|\psi_{nk}\rangle$. Figs. 2b,e show the electronic band structure of the chain isolated in vacuum obtained with considering SOI. The bands in Fig. 2e are colored with the expectation value of the $y$-component of spin, $<S_y>$, of the states. The unquenched $L_y$ of the states near the chemical potential combined with the SOI, causes a Rashba-type spin-splitting of the bands proportional to $L_y$, resulting in fully spin-polarized states in the $y$-direction. Because the states with opposite $k$ have opposite spin polarization, directional currents can be generated using circularly polarized light as demonstrated through valley-photon coupling in 2D transition metal dichalcogenides[13,14]. The calculated band gap is 100.3 meV, and the states near the chemical potential mostly derive from Te atoms, as shown in Figs. 2c,f.

Fig. 3 shows the calculated atomic and electronic structures of the segmented chain encapsulated inside an (8,8) CNT. The encapsulation does not alter the electronic structure of the segmented chain significantly except for the band repulsion at ~0.3 eV above the Fermi energy. The states near the chemical potential remain fully spin-polarized in the *y*-direction, and the energy difference between the top of the highest occupied band and the bottom of the lowest unoccupied band remains at ~100 meV, similar to that of the chain isolated in vacuum. No appreciable covalent bonding between the chain and CNT is found. Mülliken charge analysis[15] reveals that the amount of charge transferred from the CNT to the segmented chain is 0.118 $e$/Hf$_2$Te$_9$ formula unit (f.u.), where *e* is the electron charge. The binding energy of the chain, $E_b$, is calculated to be 3.32 eV/f.u., which is defined as $E_b = E_{\text{Hf2Te9}} + E_{\text{CNT}} - E_{\text{Hf2Te9/CNT}}$, where $E_{\text{Hf2Te9}}$, $E_{\text{CNT}}$ and $E_{\text{Hf2Te9/CNT}}$ are the total energies of the segmented chain isolated in vacuum, an empty CNT, and the joint system of the chain encapsulated inside the CNT, respectively.

**Topological Properties**

We now consider the topological properties of the segmented chain. Because of the presence of time-reversal symmetry and mirror symmetry with respect to $M_z$, the Zak phase of the *n*-th band, $\gamma_n$, is quantized to 0 or π (mod 2π)[16], corresponding to a topologically trivial or nontrivial band, respectively. The symmetry-protected topological invariance, $Z_2$, can be obtained by $(-1)^{Z_2} = e^{i \sum_{n \in I} \gamma_n} = e^{2\pi i P^I / e}$, where the sum of the Zak phase is over the occupied bands in only one channel $\{u^I_{nk}\}$ out of two time-reversal channels $\{u^s_{nk}\}$, s=I, II, and $P^I$ is the partial polarization over the channel I[17,18]. Since the sum of Zak phase over one channel is nothing but the partial polarization of the channel, the $Z_2$ invariance can be obtained alternatively by $Z_2 = 2P^I/e = Q$ mod 2, where $Q$ is the end charge of the finite-length chain[18,19].

We find that the topological invariance $Z_2$ of the segmented chain can be tuned by doping as shown in Fig. 4. As shown in Figs. 4a,b, in the neutral state $Z_2 = 0$ for the infinite length Hf$_2$Te$_9$ segmented chain, meaning the chain is a trivial insulator. We construct a finite length chain (~15 nm) and calculate the end charge $Q$ (see Extended Data); here $Q = 0$, confirming the topological trivialness of the undoped material[18,19]. On the other hand, the highest occupied band is nontrivial and well separated from other bands as shown in Fig. 4b. Therefore, if the system is hole-doped (h-doped) so that the chemical potential lies between the energy levels of the highest and the second highest occupied band as illustrated in Fig.

4c-d, it becomes a topological mirror insulator, provided that the Zak phases of the bands remain unchanged and that Hamiltonian of the h-doped system may be obtained adiabatically from the neutral chain without closing the band gap. That is, the number of nontrivial occupied bands in one channel changes from even to odd, and the sum of the Zak phase of the h-doped chain in one channel differs from that of a neutral chain by $\pi$, owing to the emptying out of the nontrivial highest occupied band. Hence, upon doping, $Z_2$ changes from 0 to 1, meaning the segmented chain experiences a transition from a trivial insulator to a topological mirror insulator.

We verify this topological richness with an explicit calculation of $Z_2$ invariance for an h-doped chain as shown in Figs. 4c,d (see Extended Data as well). To simulate the hole-doped system, we subtract $2e$ / f.u. from the neutral system in the DFT calculation. With the subtracted number of electrons, the topological invariance is $Z_2 = 1$, and the end charge of the finite length chain (~15nm) is $Q = 3e$. Many studies for related systems have shown that the chemical potential can be readily altered by chemical doping[20-22], electrostatic gating[6,23,24], or application of pressure[25]. This suggests that the segmented chain $Hf_2Te_9$ represents a versatile, highly unusual, and externally tunable topological material. We note that the $Hf_2Te_9$ segmented chain is, to our knowledge, the first example of a vdW-bonded crystalline material with spatial-symmetry topological invariants[26,27]. Our discovery thus expands the class of realized topological states of matter.

**Conclusion**

In conclusion, a TMC-based, vdW-bonded segmented linear chain $Hf_2Te_9$ material has been synthesized within a protective CNT cage, and its structure determined via STEM analysis and complementary DFT calculations. Theory also reveals giant spin splitting and crystal-symmetry-protected topological properties, which are promising for applications such as spintronic nanodevices and topological transistors. The segmented linear chain is a realization of a TMC in the quasi-0D limit, where 0D blocks are vdW end-to-end bonded in a linear chain. This quasi 0D material completes the family of dimensionally-reduced TMCs, where bulk $MX_2$ compounds are typically quasi-2D and $MX_3$ compounds are typically quasi-1D (M=transition metal, X=chalcogen).

**Methods**

*Materials synthesis*: The synthesis procedure is similar to that previously reported for HfTe$_3$[4]. In general, multiwalled carbon nanotubes (CNTs) (Cheap Tubes) are annealed in air at 515 °C for 20 minutes to open the end-caps before the filling step. The as-prepared CNTs (~1-4 mg) together with a stoichiometric amount of Hf powder and Te shot (~ 560 mg in total), and ~5 mg/cm$^3$ (ampoule volume) of iodine powder (transport agent) are sealed under high vacuum (10$^{-6}$ torr) in a quarter-inch quartz ampoule. The sealed ampoule is kept at 520 °C for 7 days, and then either quenched or gradually cooled to room temperature (over 9 days). Control syntheses are also performed, in which the precursor powders contained either only one material (i.e., only Hf, Te, or I$_2$), or any combination of two reagents (e.g., only Hf and Te or only Hf and I$_2$). We find no evidence of a segmented chain structure within the CNTs in these control experiments.

*Materials characterization*: The product is dispersed in isopropanol by a bath sonicator for 15 minutes, then drop-cast onto a copper grid for STEM investigation. The imaging and spectroscopy are carried out using TEAM 0.5 operating at 80 keV, with the probe convergent angle of 35 mrad and the probe current of 70 pA, at the National Center for Electron Microscopy (NCEM). The STEM simulation is implemented by using PRISM code developed by Dr. Ophus at NCEM (open-source software available online). The average-cell calculation is done by the template matching technique written in Python to increase the signal-to-noise ratio and quality of the STEM image.

*Calculation methods*: We use the generalized gradient approximation[28], norm-conserving pseudopotentials[29], and localized pseudo-atomic orbitals for the wavefunction expansion as implemented in the SIESTA code[30]. The spin-orbit interaction is considered using fully relativistic j-dependent pseudopotentials[31] in the l-dependent fully-separable nonlocal form using additional Kleinman-Bylander-type projectors[32,33]. We use 1×1×512 Monkhorst-Pack *k*-point mesh and 1000 Ry real-space mesh cut-off for all of our calculations. The van der Waals interaction is evaluated using the DFT-D2 correction[34]. Dipole corrections are included to reduce the fictitious interactions between chains generated by the periodic boundary condition in our supercell approach[35].

**Data availability:** The data sets generated and analyzed here are available from the corresponding authors on reasonable request.

**Code availability:** The code for the STEM image analysis is available from the corresponding authors upon request.

**Acknowledgement**

This work was primarily funded by the U.S. Department of Energy, Office of Science, Office of Basic Energy Sciences, Materials Sciences and Engineering Division, under Contract No. DE-AC02-05-CH11231 within the sp2-Bonded Materials Program (KC2207) which provided for synthesis of the chains, TEM structural characterization, and theoretical modeling of relaxed structure of one segment $Hf_2Te_9$. The elemental mapping work was funded by the U.S. Department of Energy, Office of Science, Office of Basic Energy Sciences, Materials Sciences and Engineering Division, under Contract No. DE-AC02-05-CH11231 within the van der Waals Heterostructures Program (KCWF16). Work at the Molecular Foundry (TEAM 0.5 characterization) was supported by the Office of Science, Office of Basic Energy Sciences, of the U.S. Department of Energy under Contract No. DE-AC02-05-CH11231. Support was also provided by the National Science Foundation under Grant No. DMR-1206512 which provided for preparation of opened nanotubes and Grant No. DMR 1926004 which provided for theoretical calculations of the electronic band structure of the segmented chain. Computational resources were provided by the DOE at Lawrence Berkeley National Laboratory's NERSC facility and the NSF through XSEDE resources at NICS.

**Author contributions**

T. P., S. O., M. L. C. and A. Z. conceived the idea. S.S. synthesized materials. T.P., S.S., B. S., J. D. C., C. S., and P. E. performed electron microscopy data acquisition and analysis. S. O. carried out density functional calculations. S. O. and M. L. C. did the theoretical analysis. M. L. C and A. Z. supervised the project. T. P., S. O., S. S., M. L. C. and A. Z. wrote the manuscript with contribution from all other authors.

**Competing interests** The authors declare no competing interests.

**Correspondence and requests for materials** should be addressed to A. Z.

**Reprints and permissions** information is available at www.nature.com/reprints

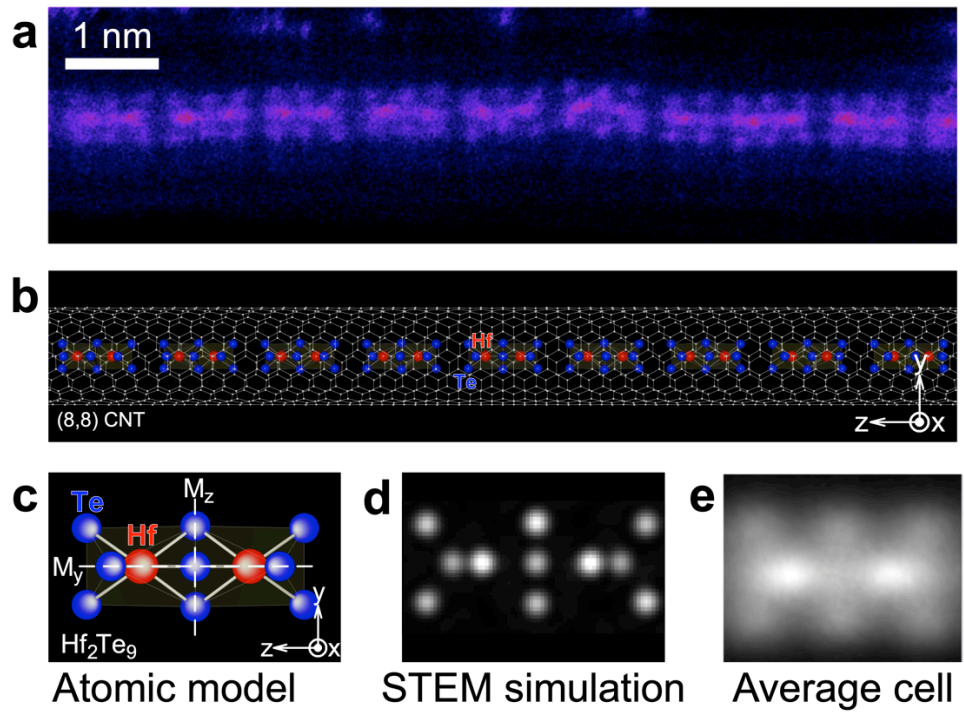

**Fig. 1. The linear segmented chain of $Hf_2Te_9$.** (a) Aberration-corrected high-angle annular dark field (HAADF) STEM image of a segmented chain of $Hf_2Te_9$ encapsulated within a double-walled CNT. A false color (Fire in the Look up Table (LUT) of ImageJ) is applied to visually aid the analysis. The structure consists of regularly-spaced segments of $Hf_2Te_9$ (or $Te_3$-Hf-$Te_3$-Hf-$Te_3$). (b-c) Atomic structure of the linear segmented chain of $Hf_2Te_9$ obtained from DFT calculations. (b) The obtained atomic structure of the segmented chain inside an (8,8) single wall CNT is shown, where Hf and Te atoms are represented as red and blue spheres, respectively, and chain direction is set to $z$-direction. (c) The building block of a $Hf_2Te_9$ unit is shown without CNT for clearer presentation of the atomic structure, in which the mirror planes perpendicular to $y$- and $z$-axes are represented by white dashed lines and denoted as $M_y$ and $M_z$, respectively. (d) Multislice simulated STEM image of a segment using the proposed atomic structure. (e) A composite STEM image generated from averaging experimentally-collected 227 orientationally-similar single segments (average cell).

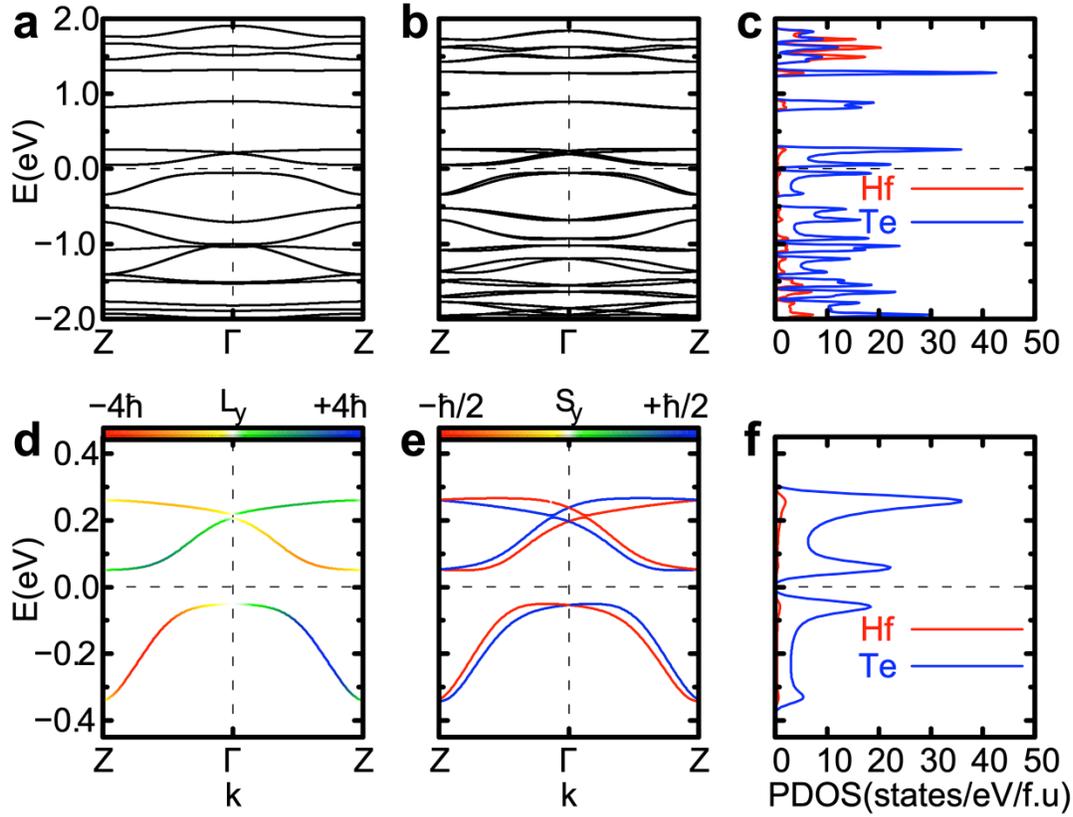

**Fig. 2**. **Calculated electronic structures of the single-atomic segmented chain of $Hf_2Te_9$**. (a-c) The electronic structure of an isolated segmented chain of $Hf_2Te_9$. The band structures obtained (a) without and (b) with considering SOI, and (c) PDOS obtained with SOI. (d-f) The electronic structure of the segmented chain $Hf_2Te_9$ near the chemical potential. The band structures obtained (d) without and (e) with considering SOI, and (f) PDOS obtained with SOI. (d-e) The color of the band lines represents the expectation value of the *y*-component of OAM <$L_y$> (in (d)), and spin <$S_y$> (in (e)). In (a-f), the chemical potential is set to zero and marked with a horizontal dashed line.

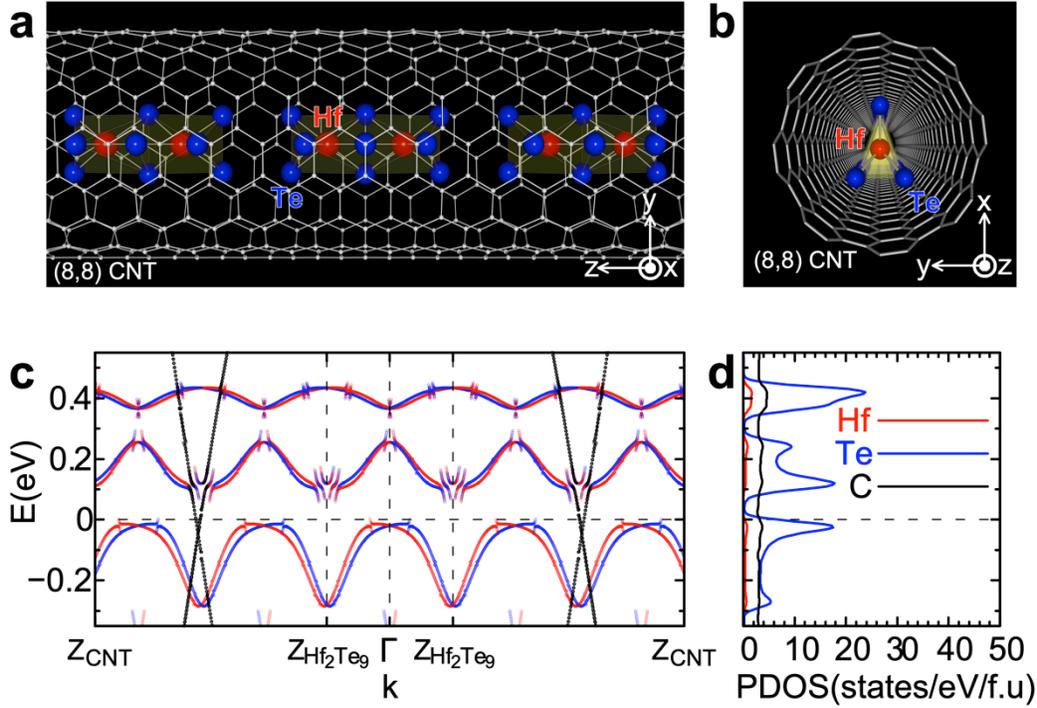

**Fig. 3. Calculated atomic and electronic structures of a linear segmented chain of $Hf_2Te_9$ encapsulated in an (8,8) CNT**. (a,b) Atomic structure of a segmented chain of $Hf_2Te_9$ encapsulated within a CNT in (a) side view and (b) axial view, where Hf, Te, and C atoms are represented as red, blue and white spheres, respectively. (c,d) Electronic structure of the segmented chain $Hf_2Te_9$ encapsulated in the CNT. (c) Band structure and (d) PDOS obtained with SOI. In (c, d), the Fermi energy is set to zero and marked with a horizontal dashed line. In (c), band structures represented by red and blue dots are projected onto spin up and spin down, respectively, in the *y*-direction of the segmented chain and unfolded with respect to the first Brillouin zone of the unit cell of the segmented chain, while the bands represented by black dots are projected onto the CNT and unfolded with respect to the first Brillouin zone of the unit cell of the CNT. The center of the Brillouin ($k = 0$) is denoted by $\Gamma$, while $Z_{Hf2Te9}$ and $Z_{CNT}$ denote the zone boundaries for the segmented chain and the CNT, respectively.

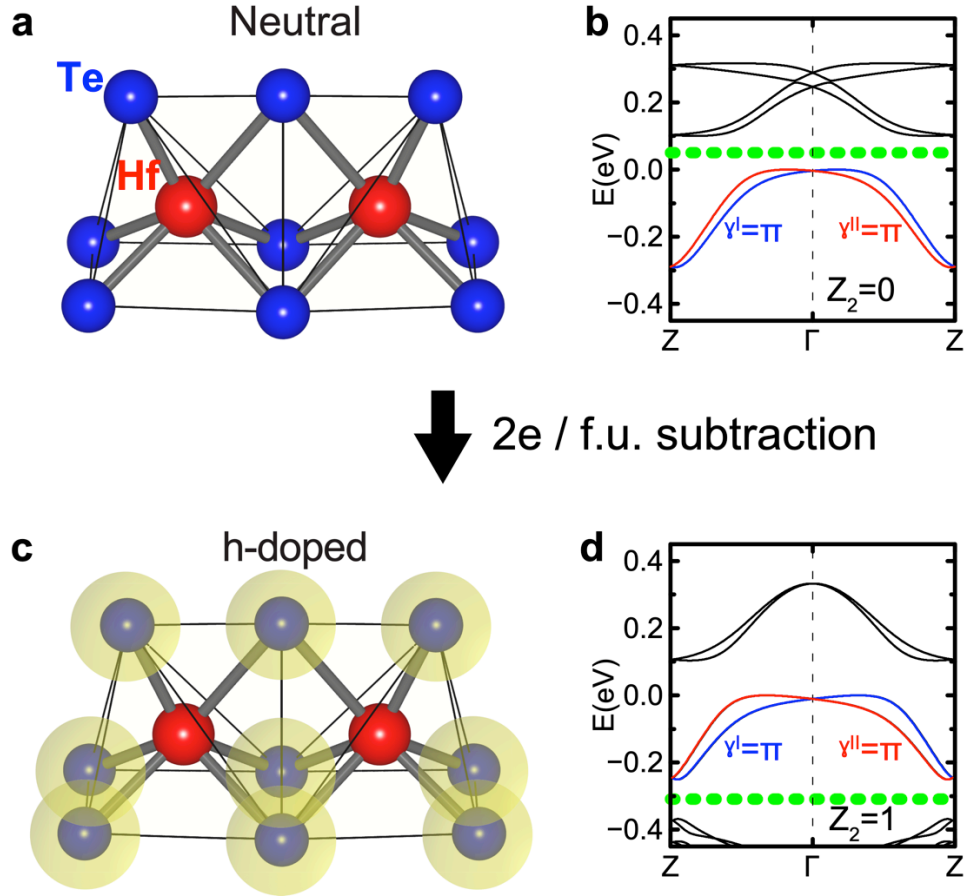

**Fig. 4. Topological properties of single-atomic segmented chain of $Hf_2Te_9$ with various chemical potential.** (a,b) The atomic and electronic structures of the segmented chain of $Hf_2Te_9$ without chemical potential change are shown for the comparison. The highest occupied band is nontrivial. (c,d) The atomic and electronic structure of the segmented chain of $Hf_2Te_9$ with $2e$ / f.u. subtracted in the DFT calculations. In (a,c), Hf and Te atoms are represented as red and blue spheres, respectively. In (c), the subtracted charge is schematically represented by yellow spheres. In (b,d), the energy level of the top of highest occupied band is set to zero for easy comparison, and the chemical potential is represented by green dashed lines. Two time-reversal channels of the highest occupied bands, s=I, II, are represented by blue and red lines, respectively. The symmetry-protected topological invariance, $Z_2$, is 0 for neutral system, and 1 for the h-doped system.

# Extended Data for Emergence of Topologically Non-trivial Spin-polarized States in a Segmented Linear Chain

Thang Pham[1,2,3,4,†], Sehoon Oh[1,2,†], Scott Stonemeyer[1,2,4,5], Brian Shevitski[1,2], Jeffrey D. Cain[1,2,4], Chengyu Song[6], Peter Ercius[6], Marvin L. Cohen[1,2], and Alex Zettl[1,2,4*]

[1]*Department of Physics, University of California at Berkeley, Berkeley, CA 94720, USA*

[2]*Materials Sciences Division, Lawrence Berkeley National Laboratory, Berkeley, CA 94720, USA*

[3]*Department of Materials Science and Engineering, University of California at Berkeley, Berkeley, CA 94720, USA*

[4]*Kavli Energy NanoSciences Institute at the University of California at Berkeley, Berkeley, CA 94720, USA*

[5]*Department of Chemistry, University of California at Berkeley, Berkeley, CA 94720, USA*

[6]*National Center for Electron Microscopy, The Molecular Foundry, One Cyclotron Road, Berkeley, CA 94720, USA*

*†These authors contributed equally*

*e-mail: azettl@berkeley.edu

**Atomic structure:**

Extended Data Fig.1 shows the atomic structure of the segmented chain $Hf_2Te_9$ obtained from DFT calculation. The atomic structure has mirror planes in the unit of $Hf_2Te_9$ denoted as $M_y$ and $M_z$, perpendicular to *y*- and *z*-axis, respectively. The length of a block is 7.89 Å, and blocks are separated by 3.65 Å. Each block connects to each other, forming a linear chain as shown in Extended Data Fig. 1(a). Extended Data Figs. 1(b-d) show the basic building block of the structure, along the *x*-, *y*-, and *z*-axis, respectively. Each block has two Hf atoms and nine Te atoms, with sequence $Te_3$-Hf-$Te_3$-Hf-$Te_3$ along the chain (*z*) direction. Each Hf atom is surrounded by six Te atoms in a trigonal prismatic form. The adjacent Te atoms form isosceles triangles. The triangle at the center of the block is on the mirror plane perpendicular to the chain direction, while those on the sides of the block cant toward the center of the block.

We would like to note that single-chain transition-metal trichalcogenides (TMTs) materials prefer either trigonal prismatic or antiprismatic structures depending on the composition, similar to the way some transition-metal dichalcogenide (TMD) materials prefer the trigonal prismatic (1H) structure while others prefer the trigonal antiprismatic (1T or 1T') structure. Therefore, both of the trigonal prismatic and antiprismatic structures are included in the candidate structures. Energetically, the segmented chain prefers a trigonal prismatic structure, unlike the continuous single-chain $HfTe_3$, which prefers an antiprismatic structure[4]. The total energy of the segmented chain in the prismatic form is ~0.8 eV lower total energy per $Hf_2Te_9$ formula unit (f.u.) than the antiprismatic form.

**Charge density:**

Extended Data Fig. 2 shows the calculated charge density of the single-atomic segmented chain of $Hf_2Te_9$. The charge density between the blocks is very low and has minimum value at the middle point between the blocks, meaning that the blocks are vdW bonded, not covalently bonded. The real-space wavefunction of all the occupied states are also examined, $\psi_{nk}(r)$, showing no occupied states between the blocks, confirming the vdW bonding of the blocks.

**Topological property:**

We calculate the Zak phases of the *n*-th bands of the channel s, $\gamma_n^S = i \int_{-\pi/a}^{\pi/a} dk \langle u_{nk}^s | \partial_k | u_{nk}^s \rangle$, for all of the occupied states of the segmented chain isolated in vacuum by integrating the Berry connection, across the 1D Brillouin zone, where $u^s{}_{nk}$ is the periodic part of the electron Bloch wave function with band index *n* and crystal momentum *k* of the channel s, and s = I, II are the time-reversal channels. Because of the presence of the time-reversal symmetry and the mirror symmetry with respect to $M_z$, $\gamma^s{}_n$ is quantized to 0 or π (mod 2π), corresponding to a topologically trivial or nontrivial band, respectively. Explicitly, We calculated $\gamma_n^S = Im[ln \prod_{i=1}^{N} \langle u_{nk_i}^s | u_{nk_{i+1}}^s \rangle]$ using the discretized *k* points, where *Im* […] is the imaginary part of […], $k_i$ with i=1,…,N, $k_{N+1}=k_1$, and a periodic gauge, in which $u_{nk_{N+1}}^s(z) = u_{nk_1}^s(z)e^{-i2\pi z/a}$, is used.

We obtain the symmetry-protected topological $Z_2$ invariance by $(-1)^{Z_2} = e^{i \sum_n \gamma_n^I}$, where the sum is over the occupied bands only in channel I. That is, the $Z_2$ invariance is determined by the number of nontrivial occupied bands in one time-reversal channel. Specifically, odd and even numbers of nontrivial occupied bands mean $Z_2$ = 0 and 1, respectively. We obtain the $Z_2$ for the neutral case and for the hole-doped case, where two electrons per formula unit are subtracted in the DFT calculation, as shown in Extended Data Fig. 3. For the neutral case, $Z_2 = 0$, corresponding to a trivial insulator, while $Z_2 = 1$ for the h-doped case, corresponding to a topological mirror insulator. The calculated Zak phase of the highest occupied band is π. As the h-doped chain has two fewer electrons per unit cell than the neutral chain, the total Zak phase of the h-doped chain must differ from that of the neutral chain by π, owing to emptying of the highest occupied band, and provided the Hamiltonian of the h-doped system may be obtained adiabatically from the neutral system without closing the band gap.

Since the sum of the Zak phase over one channel is nothing more than the partial polarization of the channel, the $Z_2$ invariance can be obtained alternatively by $Z_2 = 2P^I/e = Q$ mod 2, where *Q* is the end charge of the finite-length chain and *e* is the electron charge. We make finite chains (~15 nm) for both neutral and the h-doped system. We define the end charge of a system as the net deviation of the charge of the finite chain of the blocks on both edges from the average charge of the infinite length chain in two units. Explicitly, $Q = \rho_L + \rho_R - 2\bar{\rho}$, where $\rho_L$ ($\rho_R$) is the charge of the finite chain in the block on the left (right) edge and $\bar{\rho}$ is the charge of infinite length chain in one unit. For neutral case, $Q = 0$, meaning trivial insulator while $Q = 3e$ for the h-doped case, meaning topological mirror insulator.

**Hf$_2$Te$_9$ as a molecule:**

We investigate the stability of an isolated Hf$_2$Te$_9$ molecule (one isolated block) using DFT calculation. We construct candidate structures of the isolated molecule both in vacuum and encapsulated inside CNT using the atomic positions of the segmented chain, and the atomic positions are relaxed by minimizing total energy. Extended Data Figs. 4 and 5 show the obtained relaxed atomic structure of the isolated Hf$_2$Te$_9$ molecule in vacuum and in CNT, respectively. For both cases, the atomic structure of the molecule shows no significant change when it is isolated from the chain, except that the isosceles triangles of Te atoms on edges cant toward the center slightly more compared to the segmented chain configuration. We investigate possibilities for the molecule to change its shape to antiprismatic form, and to be separated to other small molecules and atoms. No additional possible atomic structure configurations which have lower energy than that of the structure shown in Extended Data Figs. 4 and 5 could be found.

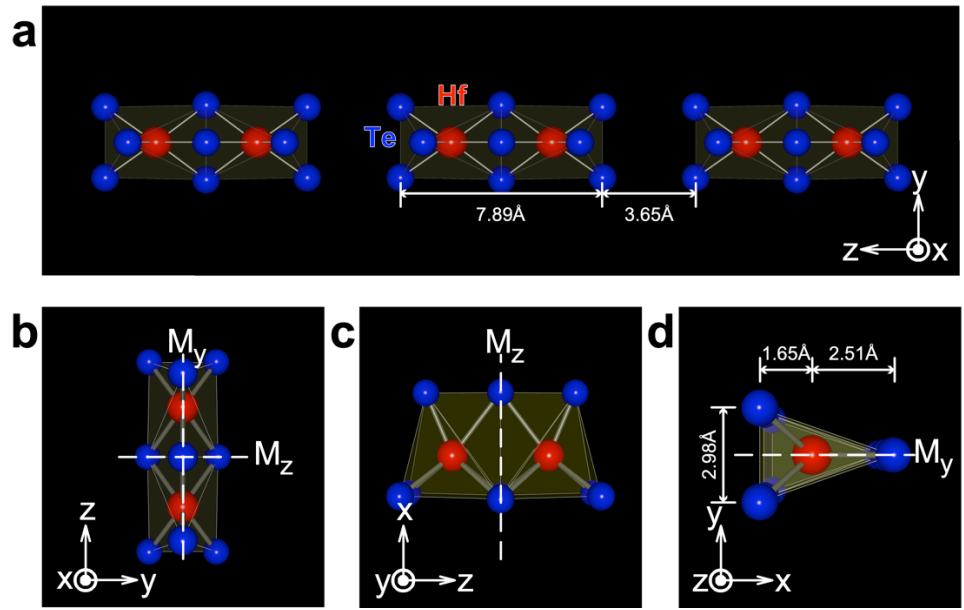

**Extended Data Fig. 1. Atomic structure of the single-atomic segmented chain of $Hf_2Te_9$ obtained from DFT calculation.** (a) The obtained atomic structure of the segmented single chain is shown, where Hf and Te atoms are represented as red and blue spheres, respectively, and chain direction is set to $z$-direction. (b-d) The atomic structure of the segmented chain of $Hf_2Te_9$ unit in side views along (b) $x$-axis and (c) $y$-axis, and (d) the axial view along $z$-axis, where the mirror planes in the unit perpendicular to $y$- and $z$-axes are represented by white dashed lines and denoted as $M_y$ and $M_z$, respectively.

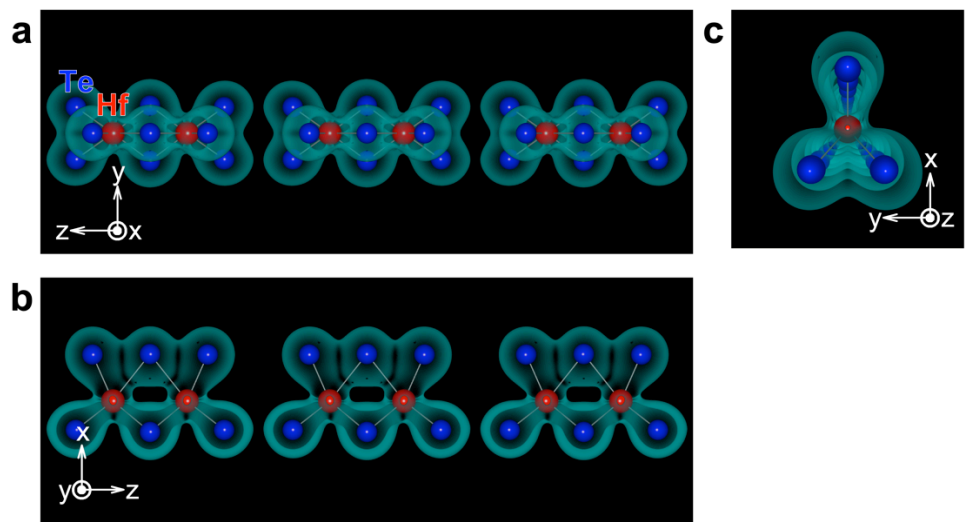

**Extended Data Fig. 2. Charge density of the single-atomic segmented chain of Hf$_2$Te$_9$.** Isosurface plots of the calculated charge density along with the atomic structures. Side views along (a) *x*-axis, (b) *y*-axis, and (c) axial view along *z*-axis. Hf and Te atoms are represented as red and blue spheres, respectively, and the isosurfaces are shaded in cyan.

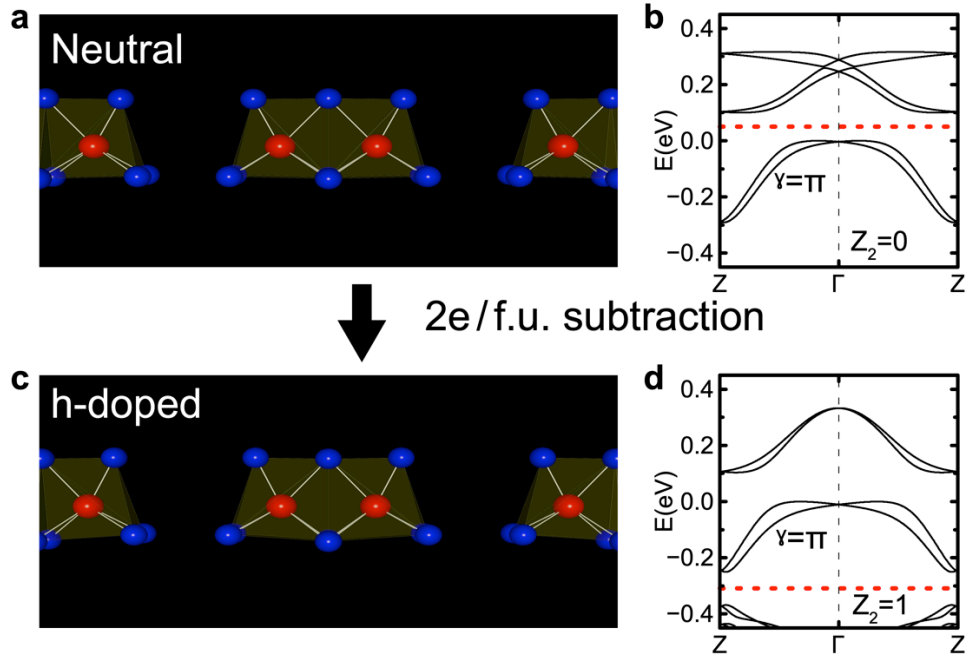

**Extended Data Fig. 3. The single-atomic segmented chain of Hf$_2$Te$_9$ with chemical potential changed.** (a, b) The atomic and electronic structure of the segmented chain of Hf$_2$Te$_9$ without chemical potential change (neutral) are shown for the comparison. (c, d) The atomic and electronic structure of the segmented chain of Hf$_2$Te$_9$ with two electron per formula unit subtracted in the DFT calculation (h-doped). In (a, c), Hf and Te atoms are represented as red and blue spheres, respectively. In (b, d), the energy level of the highest occupied band is set to zero, and the chemical potential is represented by red dashed line.

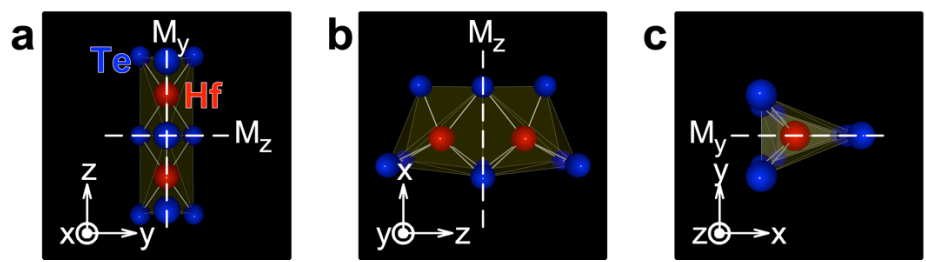

**Extended Data Fig. 4. Hf$_2$Te$_9$ molecule isolated in vacuum.** The atomic structure of the molecule is shown. Side views along (a) *x*-axis, (b) *y*-axis, and (c) axial view along *z*-axis. Hf and Te atoms are represented as red and blue spheres, respectively.

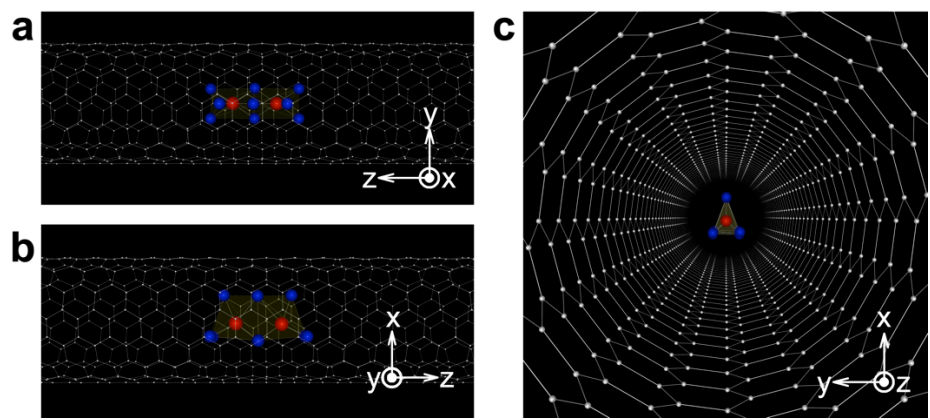

**Extended Data Fig. 5. Hf$_2$Te$_9$ molecule encapsulated inside an (8,8) CNT.** The atomic structure of the Hf$_2$Te$_9$ molecule encapsulated within the CNT is shown. Side views along (a) *x*-axis, (b) *y*-axis, and (c) axial view along *z*-axis, where Hf, Te and C atoms are represented as red, blue and white spheres, respectively.

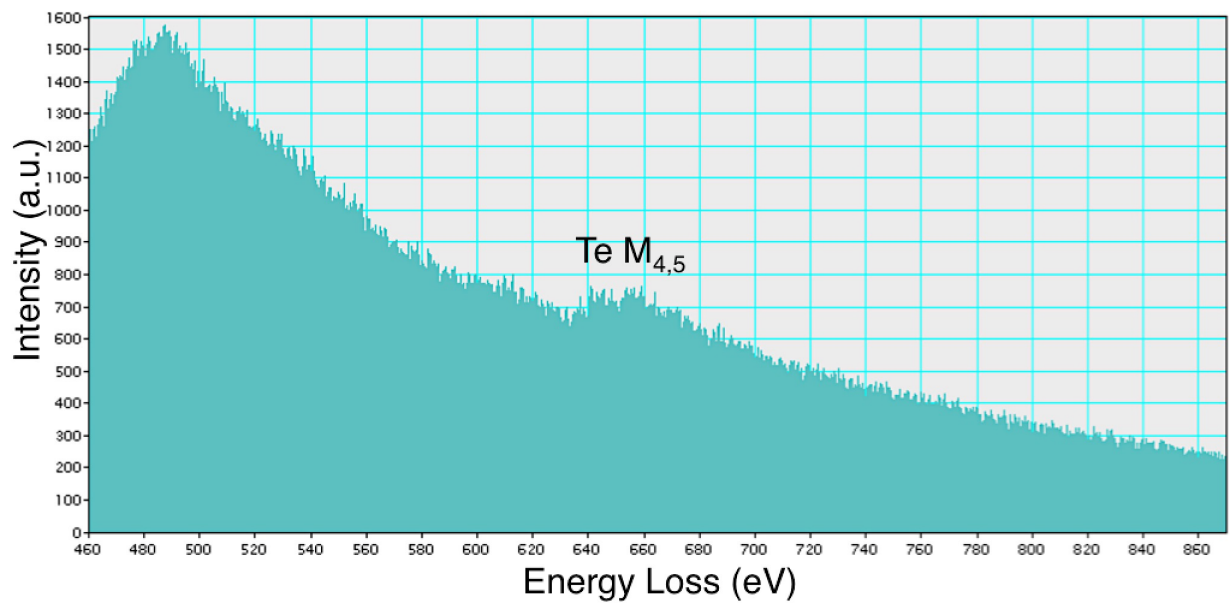

**Extended Data Fig. 6. Electron Energy Loss (EEL) spectrum of Te M edges taken from a linear chain of $Hf_2Te_9$ blocks.**